\documentclass[letter]{aa}

\usepackage{natbib}
\usepackage{graphicx}
\usepackage{subfigure}
\usepackage{amssymb}
\usepackage{txfonts}
\usepackage[colorlinks=false,bookmarks=false,pagebackref=false]{hyperref}

\hypersetup{
  pdftitle={K--H$_2$ Quasi-Molecular Satellite Detected in 
            the T-dwarf $\varepsilon$\,Indi\,Ba},
  pdfauthor={France Allard}}

\begin{document}

  \title{K--H$_2$ Quasi-molecular absorption detected in 
         the T-dwarf $\varepsilon$\,Indi\,Ba}

  \author{F. Allard
          \inst{1,2} \and N.~F. ~Allard\inst{2,3} 
          \and D.~Homeier\inst{4} 
          \and J.~Kielkopf\inst{5}
          \and M.~J.~McCaughrean\inst{6}  
          \and F.~Spiegelman\inst{7}
        }

   \offprints{fallard@ens-lyon.fr}
   \institute{Centre de Recherche Astrophysique de Lyon, 
              UMR 5574: CNRS, Universit\'e de Lyon,
              \'Ecole Normale Sup\'erieure de Lyon, 
              46 all\'ee d'Italie, F-69364 Lyon Cedex 07, France\\ 
               \and
              Institut d'Astrophysique de Paris, 
              UMR 7095: CNRS, Universit\'e Pierre et Marie Curie-Paris\,6,
              98$^{bis}$ boulevard Arago, F-75014 Paris, France\\
               \and
              Observatoire de Paris-Meudon, LERMA, UMR 8112, CNRS, 
              F-92195 Meudon Principal Cedex, France\\
              \and 
              Institut f\"ur Astrophysik, Georg-August-Universit\"at, 
              Friedrich-Hund-Platz 1, 37077 G\"ottingen, 
              Germany\\ 
               \and
              Department of Physics and Astronomy, 
              University of Louisville, Louisville, KY 40292, USA\\ 
               \and
              School of Physics, University of Exeter, 
              Stocker Road, Exeter EX4 4QL, UK\\
               \and
              Laboratoire de Chimie et Physique Quantiques, UMR5626, CNRS and 
              Universit\'e Paul Sabatier, 118 route de Narbonne 31062, 
              Toulouse Cedex, France\\ 
}

\date{Received 27 July 2007 / Accepted 27 August 2007}

\abstract{%
T-type dwarfs present a broad and shallow absorption feature centred around 
6950\,{\AA} in the blue wing of the K doublet at 0.77\,$\mu$m which 
resembles in depth and shape the satellite absorption predicted by detailed 
collisional broadening profiles.  In our previous work, the position of the 
predicted line satellite was however somewhat too blue compared to the 
observed feature. }
{In this paper, we investigate whether new calculations of the energy surfaces 
of the potentials in the K--H$_2$ system, including spin-orbit coupling, 
result in a closer coincidence of the satellite with the observed position.  
We also investigate the extent to which CaH absorption bands contribute to the 
feature {and at what $T_\mathrm{eff}$ these respective opacity
  sources predominate}. } 
{We present model atmospheres and synthetic spectra, including gravitational 
settling for an improved description of depth-dependent abundances of
refractory elements, and based on new K--H$_2$ line profiles using
improved interaction potentials.} 
{By comparison with a high signal-to-noise optical spectrum of the T1
  dwarf $\varepsilon$\,Indi\,Ba, we find that these new models do reproduce the 
observed feature, while CaH does not contribute for the atmospheric parameters 
considered. We also find that CaH is settled out so deep into the atmosphere 
that even turbulent vertical mixing would appear insufficient to bring significant 
amounts of CaH to the observable photosphere {in dwarfs of
  later type than $\sim$ L5}. }
{We conclude that previous identification of the feature at this location 
in {the spectra of T dwarfs as well as the latest L dwarfs} with
CaH was erroneous, as expected on physical grounds:  
calcium has already condensed onto grains in early L dwarfs and thus should 
have settled out of the photosphere in cooler brown dwarfs. This finding revokes 
one of the observational verifications for the cloud-clearing theory assumption:
a gradual clearing of the cloud cover in early T dwarfs.}
    

\maketitle

\section{Introduction}
Late L- and T-type brown dwarfs have atmospheres composed primarily of 
molecular hydrogen and helium.  Most refractory metals condense out to grains 
in previous evolutionary phases (early L type) and should settle below their fully 
radiative upper photospheres.  Alkali elements bind less readily to molecules or 
grains and their resonance transitions remain the last sources of opacity at
optical wavelengths, along with Rayleigh scattering by H$_2$ and He.  Alkali 
line broadening therefore defines the local ``continuum'' out to several 
thousands of {\AA}ngstr{\"o}ms from the line cores of the neutral K and Na\,D 
doublets at 0.59 and 0.77\,$\mu$m, respectively. 
{
\citet[][and references therein]{Burrows03} have shown that these
far wings deviate greatly from the simple Lorentzian shape, and
speculated on the existence of \emph{satellite features} in the blue
wings due to a local extremum in the energy shift. 
}

A number of T dwarfs show an absorption feature at 6950\,\AA{} which 
\citet{Burgasser03} identified as a CaH band.
This was a somewhat curious suggestion, however, because calcium condenses 
onto grains in L dwarfs already and is thus expected to have settled out 
of the photosphere of cooler brown dwarfs. The identification of this feature 
with CaH (as well as the observation of FeH bands) led to believe that
it was a proof for vertical mixing  
and clearing of the cloud cover in T dwarfs \citep{Burgasser03}.

\begin{figure*}
  \begin{center} 
    {\resizebox{0.98\textwidth}{!}{
        \includegraphics*[clip]{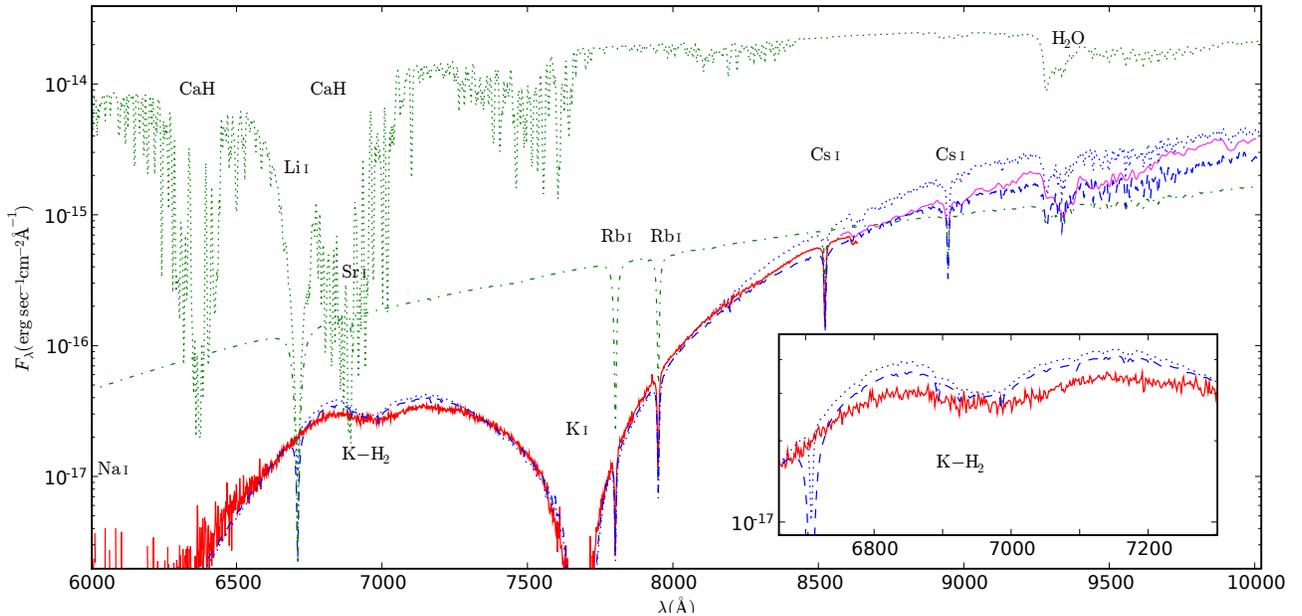} }}
\caption{FORS2 red optical spectrum of the T1 dwarf $\varepsilon$\,Indi\,Ba
(solid), compared to our synthetic spectra for a 1.3\,Gyr (dashed) 
and 2\,Gyr (dotted) model. To highlight the various sources of opacity, spectra
obtained when the K and Na\,D doublets are omitted (dot-dashed), and 
when dust grains involving Ca and other refractory species are prevented from
forming and raining out (upper dotted line) are also shown. All molecular bands 
but CaH and H$_2$O have been omitted from the latter spectrum for
clarity. 
The EFOSC spectrum of $\varepsilon$\,Indi\,Ba,b is shown from
8600\,--\,10000\,{\AA} (light [magenta] solid line) against a composite model of
both T dwarfs. 
      \label{fig:eIndi} 
    }
  \end{center}
\end{figure*}

\defcitealias{Pascale83}{P83}
\defcitealias{Rossi85}{RP85}
\defcitealias{Allard07a}{S06}
As alternative, we propose that the feature should rather be identified 
with quasi-molecular absorption by K--H$_2$. 
\cite{Allard03b} calculated 
absorption profiles of the K doublet at 0.77\,$\mu$m for H$_2$ and He 
perturbers, based upon the K--H$_2$ and K--He interaction potentials 
calculated by \citet[][hereafter \citetalias{Rossi85}]{Rossi85} and 
\citet[][hereafter \citetalias{Pascale83}]{Pascale83} using a
semi-classical approach.  
The resulting models and synthetic spectra revealed a K--H$_2$ satellite 
feature with similar depth and width to that observed at 6950\,\AA{} in the 
T dwarfs, but at a shorter wavelength of 6830\,\AA{}. In this paper, we 
present new models based on line profiles for K--H$_2$ calculated with 
updated potentials \citep[see][hereafter \citetalias{Allard07a}]{Allard07a}. 

\section{Model Atmospheres}
We have used the stellar atmosphere code \texttt{PHOENIX} \citep{jcam}
version 15.  With respect to 
previous versions used by \cite{Allard01} to determine the limiting effects 
of dust grain formation of the atmospheres of low-mass stars and brown dwarfs, 
the chemical equilibrium (CE) code in this version has been modified to 
include grain settling effects \citep{settl07}.  
For each layer of the model atmosphere, 
atarting at the deepest cloud condensation layer (as set by the CE
calculation), 
dust grain number densities are calculated in equilibrium to the gas phase.  
Condensation, gravitational settling and turbulence mixing timescale
are compared to predict the fraction of grains which must settle
\citep[see also][]{Allard03a}. 
This grain fraction is then removed from the composition and a new
equilibrium is obtained. 
This process is repeated until the grain density no longer changes. 
The advantage of this method compared to an {\em a priori\/} reduction of
refractory elemental abundances is an automatic depletion of the elemental 
abundances as a function of temperature and pressure, in the sequence of 
formation of the diverse grain types.  For example, Zr will be depleted 
first from the gas phase, as ZrO$_2$ is the grain that forms at the highest 
temperature, followed at lower temperatures by Fe, Mg (MgSiO$_3$), Al 
(Al$_2$O$_3$), and so on. The \cite{Allard01} ``Cond'' models --- which used 
an equilibrium dust distribution but simulated gravitational settling by 
commenting out dust opacities -- tended to retain some trace of the molecular
compounds involved in the grains.  Here the depletion is complete and, for 
example, no TiO absorption band opacity remains in layers where grains 
involving Ti have been found to condense.  This formalism corresponds
therefore to a more accurate description of grain settling. 
 
In the {Settl} models described above, gravitational settling
efficiently clears the uppermost region of the photosphere from dust
down to a few pressure scale heights from the top of the convection
zone. 
A full description of these models exceeds the scope of this paper
and will be published separately. 

\section{New line profiles}\label{new}
An important improvement with respect to the 2001 Cond models 
is the inclusion of the pressure broadening profiles for 
the neutral alkali resonance lines of 
{
Li\,\textsc{i} by \citet{Allard05},
Rb\,\textsc{i} and  
Cs\,\textsc{i} by \citet{Allard06}, 
K~\,\textsc{i} and  
Na\,\textsc{i}\,D 
}
as in \cite{Allard03b} 
and by \cite{Allard07a}.  These profiles include fine structure and 
use damping constants for the Lorentz line cores from semi-classical 
theory \citep{Allard07b}.

The most important contributions to the satellite feature under consideration
come from H$_2$ and He. The profiles of K-H$_2$ account for two different orientations of the 
H$_2$ molecule: $C_{2v}$, where the H$_2$ molecular axis is perpendicular to 
the K radiating atom, and $C_{\infty v}$, where all atoms are collinear.  
The $C_{2v}$ \citetalias{Allard07a} satellite is predicted to lie at 6980\,{\AA}, while the 
weaker $C_{\infty v}$ contribution peaks at 6920\,{\AA}: these should be
compared to 6851 and 6695\,{\AA}, respectively, for the \citetalias{Rossi85} profiles.  
The K--He satellite is predicted to be at 6930\,{\AA}.  Note, however, that 
when accounting for intermediate orientations of the H$_2$ molecule, 
additional satellite contributions may wash out and slightly change the 
predicted position.  \cite{Santra05} have published {\em ab initio\/} 
potentials for K--H$_2$ at 4 orientations of the H$_2$ molecule, as well as
for K-He, predicting K--H$_2$ satellites around 6920\,{\AA} and 6850\,\AA{} 
for the $C_{2v}$ and $C_{\infty v}$ orientations, respectively, with K--He 
at 7000\,{\AA}.  These values correspond to the extrema of the difference 
potentials, however: our corresponding values are 6906\,{\AA} and 6830\,{\AA} 
for C$_{2v}$ and $C_{\infty v}$, respectively. Finally, \citet{Zhu06} have 
also computed profiles of K and Na perturbed by He, finding a K-He satellite 
at 7080\,{\AA}. Combined, these calculations predict a broad absorption 
feature with properties very similar to that observed at 6850 to 7100\,{\AA} 
and centred at 6950\,{\AA}) in the spectra of T dwarfs such as 
$\varepsilon$\,Indi\,Ba as discussed below. 

\section{Optical spectral analysis of the T1 dwarf $\varepsilon$\,Indi\,Ba}

The {$\varepsilon$\,Indi\,Ba,b} binary system is a unique test
for brown dwarf models. 
They are the closest known brown dwarfs to the Sun, allowing
spectra to be obtained with high signal-to-noise in the faint optical
range. 
The well-determined HIPPARCOS distance (3.626\,pc) allows 
accurate comparison to be made with models at an absolute level.  
Based on their $J\!H\!K$ photometry and $H$-band spectra,
{\cite{IndiBab04}} obtained effective temperatures of 1276\,K and 854\,K for 
Ba and Bb, respectively, using the \cite{Baraffe03} 1.3\,Gyr Cond isochrone 
for the most likely age of the $\varepsilon$\,Indi\,AB system.
This corresponds to surface gravities of $\log g$\,=\,5.15 and 4.9, and radii 
of 0.091 and 0.096\,R$_\odot$, respectively.  
{
We have compared a {Settl} model calculated for these parameters to a 
medium-resolution resolved optical spectrum of the T1 primary (Ba), as
shown in Fig.~\ref{fig:eIndi}.
}

\begin{figure}
  \begin{center}
   \includegraphics[width=0.48\textwidth,clip]{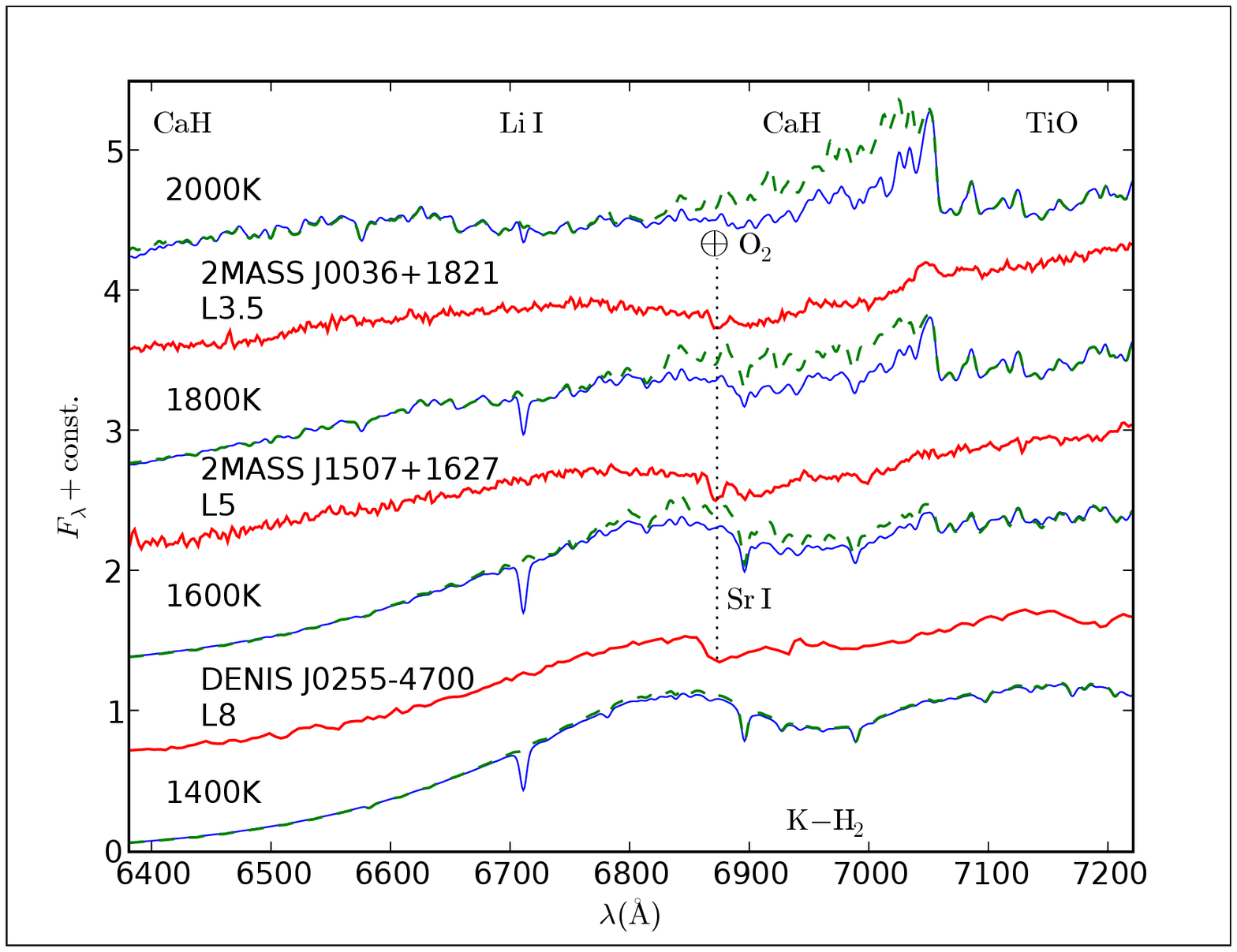}
    \caption{Effective temperature sequence of {log\,$g$\,=\,5.0}
      Settl models with (solid) and  
    without (dashed) CaH band opacity. The synthetic spectra have been 
    degraded to 5\,{\AA} resolution for comparison with L dwarf 
    spectra (thick [red] lines) observed with the Keck LRIS spectrograph
    \citep{2001AJ....121.1710R,2000AJ....119..369R,1999AJ....118.2466M}. 
    {
    Spectral subtypes of the L dwarfs are given in the optical
    classification system of 
    \citet[in prep.\protect\footnotemark]{krlLdwarf99}; 
    note that {DENIS~J\,0255$-$4700} has been typed L6 in the system of 
    \citet{1999AJ....118.2466M}. 
    }
    The sharp CaH band head near 7075\,\AA{} vanishes in 
    both observed and model spectra between about L5 and L8, and
    a distinctive absorption feature emerges that extends 
    well to the red of the original CaH band head. 
    The models have Li abundance 
    reduced to 1\,\% (solid) and 0.1\,\% (dashed), 
    indicating strong Li depletion in all these field L dwarfs. 
    \label{fig:CaHvsTeff}
    }  
  \end{center}
\end{figure}

As part of a comprehensive 0.6--5$\mu$m photometric and spectroscopic study
of $\varepsilon$\,Indi\,Ba,b, the ESO VLT was used on June 16 2004 to obtain
optical (0.6--0.86$\mu$m) spectroscopy using FORS2 in long-slit mode, with a
slit width of 0.5 arcsec and the HR collimator. In this mode, the 600RI grism
yielded a spectral resolution of R$\sim$1000 at 6780\,\AA, while the seeing of
0.35 arcsec FWHM ensured that the two T dwarfs, separated by $\sim$0.8 arcsec
at that epoch, were well-resolved spatially. A total integration time of 80 min
was obtained by co-adding $6\times 800$\,s individual exposures. The data
reduction was standard, using skylines for wavelength calibration, dome lights
for spectral flat-fielding, and the DC white dwarf spectrophotometric standard,
LTT\,9491, for flux calibration. Unfortunately, no direct telluric calibration
was possible, but an excellent cancellation of all known telluric features was
achieved by using the NSO Kitt Peak atmospheric transmission spectrum
\citep{hinkle03} convolved to the FORS2 resolution and with iterative
adjustment of the effective airmass to match the conditions over Paranal at
the time of our observations. 
{
This allowed efficient removal of the 
6884\,{\AA} O$_2$\,B band close to the satellite (compare
Fig.~\ref{fig:CaHvsTeff}). 
}
Further details of the data reduction will be
given by King et al.\ \citetext{2007, in prep}.
Preliminary absolute flux calibration was obtained by using the combined
$I$-band photometry from the DENIS second data release
\footnote{\href{http://vizier.u-strasbg.fr/viz-bin/Cat?B/denis}{http://vizier.u-strasbg.fr/viz-bin/Cat?B/denis}}, 
as listed in \citet{IndiBab04}, to calibrate 
{
a composite spectrum of $\varepsilon$\,Indi\,Ba,b taken with the EFOSC
spectrograph at the ESO 3.6\,m telescope, the red part of which 
} 
is also shown in
Fig.~\ref{fig:eIndi}. The FORS2 spectrum was then rescaled to match
the composite spectrum at $\lambda < 8000$\,{\AA}, where the flux
contribution of the Bb component is less than 1\,\%.  

\addtocounter{footnote}{-1}  
\addtocounter{footnote}{-1}
\stepcounter{footnote}\footnotetext{\href{http://DwarfArchives.org}{see http://DwarfArchives.org}}

Our synthetic spectrum reproduces the near-IR photometry for Ba to
within 0.15\,mag, supporting our treatment of dust opacity, which 
shapes the infrared SED in transition objects.  
The red optical region is extremely well reproduced in most
details, including the shape of the K doublet at 0.77\,$\mu$m and the
depth of the secondary alkali lines of Rb\,\textsc{i} and Cs\,\textsc{i},
absorbing through the red wing of the K\,\textsc{i} doublet. 
Assuming a meteoritic, i.\,e.\ proto-solar, lithium abundance, the
model also predicts a strong Li\,\textsc{i} 6707\,\AA{} 
doublet, but this is not observed.  This implies a relatively massive
object which depleted its Li at a young age when its atmosphere
was fully convective and its core temperature exceeded the Li burning 
threshold: this process {consumes most Li} in brown dwarfs 
more massive than 0.055\,--\,0.06\,M$_\odot$ in the first Gyr of
their lifetime \citep{2005AN....326..948Z}. 

According to the \citet{Baraffe03} evolutionary tracks, a
0.055\,M$_\odot$ brown dwarf at an age of 2.0\,Gyr should have
$T_\mathrm{eff}$\,=\,1350\,K, $\log g$\,=\,5.30 and
$R$\,=\,0.086\,R$_\odot$. This model is 
overplotted as dotted line in Fig.~\ref{fig:eIndi} and still agrees
with the data within observational and modelling uncertainties. 
For a qualitative estimate of the degree of lithium depletion we have
reduced the elemental Li abundance in this latter model by 
1.5\,dex. 
{
It still shows a clearly visible resonance line not
detectable in the observation, confirming very low Li abundance. 
}

The models also show a weak 6894.5\,{\AA} Sr\,\textsc{i} line 
against the background of the alkali lines. 
It is not detectable in the observed spectrum, 
indicating incomplete treatment of Sr
condensation. 

\begin{figure}
   \begin{center}
  \includegraphics[width=0.48\textwidth,clip]{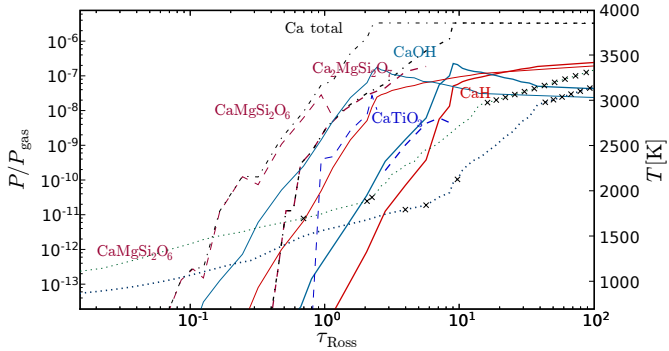}
   \caption{Partial pressures of principal Ca-bearing species versus optical 
depth in the Settl model for $T_\mathrm{eff} = 1280$\,K (thick lines) and  
1600\,K (thin lines). Gas temperature is shown (dotted lines) with the 
convection zones indicated (crosses), right scale. 
      \label{fig:ppCa}
    }  
  \end{center}
\end{figure}

\subsection{The K--H$_2$ satellite}
\label{sec:kh2sat}
The key feature in this high signal-to-noise spectrum for present purposes is 
the shallow absorption feature centred at 6950\,{\AA}, the exact location of 
the satellite of the K--H$_2$ interaction with the 0.77\,$\mu$m doublet as
newly predicted here using the \citetalias{Allard07a} profiles (Figs.~\ref{fig:eIndi} and 
\ref{fig:CaHvsTeff}). However, the new satellite apparently extends
not far enough to the red to 
reproduce the more elongated shape of the observed feature. 
K--He produces a satellite which might reproduce the observed extent of 
the feature to 7100\,{\AA} according to the \cite{Santra05} and \cite{Zhu06} 
calculations.  Our old potentials predict it further to the blue, from 
6930--6980\,{\AA}, thus superimposing on the location of the 
K--H$_2$ satellite and producing an absorption trough that is too narrow. 
         
\subsection{The CaH $A^2{\pi}$--$X^2{\Sigma}$ band}
The {dot-dashed line} in Fig.~\ref{fig:eIndi} shows
the effect of eliminating the large wings of the K and Na\,D doublet,
to reveal other opacity sources which participate in the
formation of the spectrum. 
{
The Settl model does not predict any features besides the
Sr\,\textsc{i} line (compare also Fig.~\ref{fig:CaHvsTeff}) 
in the spectral region of the satellite. 
}

However, if we artificially prevent calcium from being locked into grains and 
settled out, the $A^2{\pi}-X^2{\Sigma}$ band of CaH is seen, partly 
coincident with the observed feature (upper dotted line in
Fig.~\ref{fig:eIndi}). 
This led \citet{Burgasser03} to propose that CaH is responsible
for the observed feature in {early to mid-T dwarfs}, advocating
a possible resurgence of CaH due to upwelling of
CaH or cloud-clearing effects.   

We find that most calcium is locked into diopside (CaMgSi$_2$O$_6$) and 
akermanite (Ca$_2$MgSi$_2$O$_7$)  below 1900\,K, leading to a
depletion of calcium to 10$^{-4}$ or less as these 
condensates settle out (Fig.~\ref{fig:ppCa}). CaH only prevails at solar 
abundances deep in the optically-thick atmosphere, independent of whether 
we account for clouds or not. 

Fig.~\ref{fig:CaHvsTeff} illustrates the difference between models with 
and without CaH bands for effective temperatures ranging across the late-L 
sequence. At 2000\,K, the CaH band does indeed shape the pseudo-continuum at 
6800--7050\,\AA, but by 1600\,K, the K--H$_2$ satellite is dominant and 
CaH barely contributes to the absorption. 
{
 While the intensity of the satellite actually increases with gas
 temperature \citepalias[see Figs. 10-12 of][]{Allard07a}, in 
 our model spectra the satellite grows in strength with 
 \emph{decreasing} $T_\mathrm{eff}$, due to clearing of the atmosphere 
 and increased pressure in the line-forming region 
 \citep[see][for details]{settl07}. 
It creates an absorption trough extending from 6850 to 7100\,{\AA}, 
visible in the observed spectra of the late L dwarfs, where it is 
extending beyond the CaH band head and even further to the red than in
the models (cf.\ \S\,\ref{sec:kh2sat}).  
Once T dwarf temperatures are reached, our models retain essentially
no CaH opacity, challenging the cloud-clearing picture 
unless condensable gas is upwelled efficiently over more than two pressure
scale-heights by turbulence. 
}

\section{Summary}
We have computed new model atmospheres based on detailed absorption 
profiles for the neutral Na\,D, Li, K, Rb, and Cs alkali lines (\citetalias{Rossi85} and \citetalias{Pascale83} 
potentials) and, in particular, new interaction potentials \citepalias{Allard07a} for 
K--H$_2$. These new models predict a K--H$_2$ satellite absorption 
feature at 6950\,\AA{} closely matching the position and shape of an observed 
feature in the spectrum of the T1 dwarf $\varepsilon$\,{Indi}\,Ba. 
We therefore conclude that the K--H$_2$ satellite is the most natural 
explanation for the feature, rather than CaH as has been previously
proposed. Indeed, the high (solar) abundance of CaH which would be required
to generate absorption near this spectral location should only be found deep
in the optically-thick atmosphere at the $T_\mathrm{eff}$=1280\,K of
$\varepsilon$\,Indi\,Ba, preventing its bands from appearing in spectra even 
assuming a clearing of the surface cloud coverage.      

\begin{acknowledgements}
Part of this work was financially supported by the PNPS program of the CNRS
and the EC MC RTN CONSTELLATION (MRTN--CT--2006--035890). We also thank the 
CINES, IDRIS, and the GWDG 
for generous allocations of computing time necessary to complete 
this project, 
and an anonymous referee for valuable comments on the manuscript. 
NFA is grateful to A.\ Staiano for his help in computing K--H$_2$ opacity
tables and 
DH to Sandy Leggett for electronic versions of the LRIS spectra. 
This work has also benefited from the M, L, and 
T dwarf compendium housed at \href{http://DwarfArchives.org}{DwarfArchives.org} and 
maintained by Chris Gelino, Davy Kirkpatrick, and Adam Burgasser, and from the 
IAC ultracool dwarf catalogue at
\href{http://www.iac.es/galeria/ege/catalogo\_espectral}
     {www.iac.es/galeria/ege/catalogo\_espectral}
compiled by Juan Cabrera and Elena Cenizo under the direction of
Eduardo Mart{\'{\i}}n. 
\end{acknowledgements}

\end{document}